\documentclass[5p]{elsarticle}

\usepackage{hyperref}
\usepackage{gensymb}
\usepackage{graphicx}
\usepackage{subcaption}
\usepackage{url,array}
\usepackage{color,soul,xcolor}
\usepackage{textcomp}
\usepackage{longtable}

\journal{Physics Letters A}
	
\begin{document}

\begin{frontmatter}

\title{Hard Ferromagnets as a new Perspective on Materials for Thermomagnetic Power Generation Cycles}

\author[gradcenter,brooklyn]{Anthony N. Tantillo}
\author[vac]{Alexander Barcza}
\author[vac]{Volker Zellmann}
\author[inrim]{Vittorio Basso}
\author[paris]{Martino LoBue}
\author[neel]{Nora M. Dempsey}
\author[gradcenter,brooklyn]{Karl G. Sandeman\corref{mycorrespondingauthor}}
\cortext[mycorrespondingauthor]{Corresponding author}
\ead{karlsandeman@brooklyn.cuny.edu}

\address[gradcenter]{Physics Program, The Graduate Center, CUNY, 365 Fifth Avenue, New York 10016, USA}
\address[brooklyn]{Department of Physics, Brooklyn College, 2900 Bedford Avenue, Brooklyn 11210, USA}
\address[vac]{Vacuumschmelze GmbH \& Co. KG, Gr\"{u}ner Weg 37, 63450, Hanau, Germany}
\address[neel]{Univ. Grenoble Alpes, CNRS, Grenoble INP, Institut N\'{e}el, 38000 Grenoble, France}
\address[inrim]{Istituto Nazionale di Ricerca Metrologica, Strada delle Cacce 91,10135 Torino, Italy}
\address[paris]{Universit\'{e} Paris-Saclay, ENS Paris-Saclay, CNRS, SATIE, 91190, Gif-sur-Yvette, France}

\begin{abstract}
We consider the ways in which magnetically hard materials can be used as the working materials in thermomagnetic power generation (TMG) cycles in order to expand the area in the magnetisation vs. applied field ($M-H$) plane available for energy conversion. There are 3 parts to this Perspective. First, experiments on commercially available hard ferrites reveal that, while these materials are not yet good TMG candidates, hard ferromagnets with higher thermal conductivity and a greater change of magnetization with temperature could outperform existing TMG materials. Second, computational results indicate that biasing a soft magnet with a hard ferromagnet is essentially equivalent to shifting the $M-H$ loop by an amount proportional to the field of the biasing magnet. Work outputs under biased conditions show a substantial improvement over unbiased cycles, but experimental verification is needed. Third, we discuss the rationale for exploring artificial spin reorientation materials as novel TMG working materials.
\end{abstract}


\end{frontmatter}

\section{Introduction}
\subsection{Energy harvesting and pyromagnetic heat engines}
The industrial ability to generate power from heat is well established, but it suffers from a lack of widespread adoption, despite the magnitude of available waste heat sources~\mbox{\cite{Chan2013,Xu2019}}.  In 2017, industrial energy use accounted for 37{\%} of the 9.7~gigatonne of oil equivalent (GToe) worldwide total final consumption{~\cite{IEA_2019}}.   Much of the energy consumed by industry is in thermal processes such as cement-making and steelmaking and about 20-50{\%} of this is discarded as heat{~\cite{EERE_2004}}.

Many commercial heat energy conversion devices employ an indirect approach, first converting heat to mechanical work, and then using this mechanical energy to drive an electric engine.  The most commonplace method is the organic Rankine cycle (ORC), which has an energy generation cost of \$1500-3500/kW~\cite{cunningham_2002}. However, the relatively low efficiency of the ORC in small-power (kW-scale) settings means that alternative working fluids and cycles are being sought.   This search is complicated by the fact that solid-state alternatives to fluid-based heat-to-mechanical indirect energy generation are few, relatively expensive, and are currently inefficient.  Thermoelectric power generation (TEG) is arguably the most developed, and is based on the Seebeck effect - the generation of a voltage across a material by a temperature difference.  Studies have concluded that the optimal deployment of TEGs is not currently in the low-grade heat environments targeted by the materials studied here~\cite{Kishore2018a,Thekdi2021}.

This article aims to provide a novel perspective on research into pyromagnetic effects (PMEs), and in doing so, open up a new source of magnetically hard, anisotropic power conversion materials.  The pyromagnetic effect is the induction of a magnetization by the action of a change of temperature.  It is strong around a magnetic phase transition and is the corollary of the more often-studied magnetocaloric effect, the basis for magnetic field-driven cooling~\cite{sandeman_2021}.  Both Edison, in 1888~\cite{Edison_1888} and Tesla (in 1889~\cite{Tesla_1889} and 1890~\cite{Tesla_1890}) published the first patents on the use of iron near its Curie point as a working material for conversion of heat into electricity but a lack of suitable pyromagnets to cover a range of applications has arguably resulted in only a sporadic research literature until now.

Interest in energy harvesting, combined with the discovery of new magnetocaloric materials that have been proposed as pyromagnets, has resulted in increased attention for the field of TMG~\cite{Srivastava_2011,hsu,Christiaanse2014,Almanza2017a,Chun2017,Gueltig2017,Waske2019,Dzekan2021,Dzekan2021a,Ahmim_2021b}.   One method for using the pyromagnetic effect in TMG is to surround a heated ferromagnet with a coil.  When the ferromagnet becomes paramagnetic above its Curie temperature, the changing flux in the external coil results in a current~\cite{Tesla_1890}.  This is direct conversion to electricity of the energy associated with the magnetic field of the pyromagnet. By contrast, indirect methods such as a Curie wheel are also feasible~\cite{Bremer_1904}.  In a Curie wheel, the active pyromagnetic material is arranged on the circumference of a wheel that is rotated in a variable magnetic field.  During rotation, the pyromagnet is heated and cooled by external fluids.  The hot fluid is essentially the energy source in this scenario.  Cold parts of the pyromagnet ring are more attracted to the high field zones, resulting in a torque and thereby conversion of heat into mechanical energy~\cite{Bremer_1904}.   A review of the history of thermomagnetic devices can be found in Refs.~\cite{Kishore_2018,Kitanovski_2020}.  Recent commercial activity in this area includes a Curie-wheel motor which used gadolinium with a minimum temperature span of 20~\degree{C} to generate power in the kilowatt range with zero emissions~\cite{swiss_blue}.

As discussed by Hsu et al.~\cite{hsu}, the measure of efficiency, $\eta$, of a pyromagnetic heat engine is the ratio of work output ($W$) to heat input ($Q_h$) and has a Carnot limit, $\eta_{\rm{Carnot}}$, defined by the temperature of the heat source, $T_{\rm{hot}}$ and that of the heat sink,  $T_{\rm{cold}}$:  
\begin{equation}
\eta={W \over Q_h}
\label{eff_eq}
\end{equation}

\begin{equation}
\eta_{\rm{Carnot}}=1-{T_{\rm{cold}} \over T_{\rm{hot}}}
\label{eq:Carnot}
\end{equation}

\begin{equation}
\eta_{\rm{rel}}={\eta \over \eta_{\rm{Carnot}}}
\end{equation}
	Here, $\eta_{\rm{rel}}$ is defined as the engine efficiency relative to that of the Carnot limit.  Low grade waste heat (which the US Department of Energy defines as heat supplied at temperatures below 230{\degree}C ~\cite{BCSInc2008}) is sufficiently abundant as to make it the prime target for technology development, despite its reduced Carnot efficiency according to Equation~\ref{eq:Carnot}. Pyromagnetic heat engines are expected to function at up to 55\% of Carnot efficiency at temperatures a few hundred degrees Kelvin above room temperature~\cite{hsu,Brillouin_1948}.  By contrast, the efficiency of typical TEGs has not exceeded 20\% of the Carnot value~\cite{Kishore2018a}.   

\subsection{The role of pyromagnetic material properties}
	The efficiency of a pyromagnetic heat engine is ultimately the property of a given device. However, the maximum efficiency of any device can be derived from material properties, depending on the thermodynamic cycle employed. The starting point for determining the potential pyromagnetic output of a magnet is the examination of a thermomagnetic cycle of the ``working material" such as the one shown schematically in Figure~\ref{fig:schematic_cycles}(a).   The magnetization, $M$, of a conventional ferromagnetic pyromagnet will decrease as a function of temperature ($\partial M / \partial T < 0$).  It is also possible for an unconventional pyromagnet, in which $\partial M / \partial T$ is instead positive, to be used and the thermodynamic cycle to be reversed. However, we will focus on the more common conventional effects.
	
	In Figure~\ref{fig:schematic_cycles}(a), the isothermal magnetisation of the pyromagnet is shown as a function of magnetic field at two temperatures, $T_{\rm{hot}}$  and  $T_{\rm{cold}}$, which are those of the heat source and heat sink, respectively.  In the example cycle, the material begins at point (1) in a low temperature, high magnetization state at a low applied field value. The magnetic field is increased isothermally up to point (2). Then, the material is heated at a constant magnetic field, dropping its magnetization to point (3). From point (3) to point (4), the field is decreased at constant temperature, bringing the material to a low magnetization, high temperature state.  The last step in the cycle is to cool the material in a fixed field, to return to the starting position (1).  
	The work output, $W$ from the pyromagnetic working material is given by the area bounded by the magnetic isotherms at $T_{\rm{hot}}$ and at $T_{\rm{cold}}$\cite{Solomon_1988}:  
\begin{equation}
W = -\mu_0 \oint H \,dM = \mu_0 \oint M \,dH,
\label{work_eq}
\end{equation}
where $H$ is the magnetic field applied.  This area (shaded in the $M$ vs $H$ plots in Figures~\ref{fig:schematic_cycles}(a) and (b)) should be maximized.  

While it is not a focus of our Perspective, we complete this subsection with a statement on the cycle efficiency.  From Equation~\ref{eff_eq}, we need to calculate the heat input to the engine, $Q_h$.  From conservation of energy:
\begin{equation}
Q_h=Q_c + W \, ,
\label{total_eq}
\end{equation}
where $Q_c$ is the heat rejected by the engine to the cold sink.  As explained by Solomon, the $Q_c$ term comprises two contributions, when heat flows into the cold sink.  In our cycle, these are in steps (4)$\rightarrow$(1) and (1)$\rightarrow$(2) .  During step (1)$\rightarrow$(2), an amount of heat $T_{\rm{cold}} \Delta S$ flows into the cold sink, resulting from changing the magnetic field applied to the working material at fixed temperature $T_{\rm{cold}}$.  During step (4)$\rightarrow$(1), the working material is cooled from $T_{\rm{hot}}$ to $T_{\rm{cold}}$ in a finite field. Therefore, the amount of heat passed to the cold sink during that step is $\int_{T_{\rm{cold}}}^{T_{\rm{hot}}} C_p(T,H_1) dT$ where $C_p(T,H_1)$ is the heat capacity of the active pyromagnetic material at constant pressure and magnetic field $H_1$.  In this nomenclature, we may write the efficiency of a pyromagnetic engine as:

\begin{equation}
\eta= { \mu_0 \oint M \,dH     \over    \int_{T_{\rm{cold}}}^{T_{\rm{hot}}} C_p(T,H_1) dT + T_{\rm{cold}}\Delta S + \mu_0 \oint M \,dH}.
\label{efficiency_equation}
\end{equation}

\onecolumn
\begin{small}
\setlength\LTleft{0pt} 
\setlength\LTright{0pt} 
\begin{longtable}{| p{0.04\linewidth} | >{\centering} p{0.08\linewidth} | >{\centering}p{0.11\linewidth}|>{\centering}p{0.08\linewidth}| >{\centering}p{0.07\linewidth}|>{\raggedright\arraybackslash} p{0.5\linewidth}|}
\hline
\textbf{Year} & \textbf{Ref.} & \textbf{Working material} & \textbf{Field source}  &\textbf{Type}& \textbf{Comments} \\
\hline \hline
\hline1884&\cite{McGee_1884} (A)&Iron&Perm.&P.R.&First description of a Curie wheel as a ``novel magnetic engine".\\
\hline1887&\cite{Edison_1887} (A)&Iron&Either&P.R. + Static&Edison's first article on TMG, describing a rotating armature powered by heat and a means of converting that rotation into electricity.\\
\hline1888&\cite{Edison_1888} (P)&Iron&Either&P.R.&First TMG patent, based on kinetic output from a rotating armature (an aspect of 1887 article).\\
\hline1889&\cite{Stefan_1889} (P)&Nickel&Perm.&P.R.&Demonstrator made for educational purposes, using a nickel arc on a pendulum. First research article to provide a thermodynamic analysis.\\
\hline1889&\cite{Tesla_1889} (P)&Iron&Either&Linear&Functional armature used to harvest energy.\\
\hline1890&\cite{Tesla_1890} (P)&Iron&Perm.&Static (1)&First patent of an induction device.\\
\hline1892&\cite{Edison_1892} (P)&Iron&Either&P.R. + Static (3+)&Device based on second half of 1887 paper, designed to produce dc or alternating current via induction-based energy recovery.\\
\hline1904&\cite{Bremer_1904} (P)&Unspecified&E/M&P.R.&First patent of a Curie wheel.\\
\hline1930&\cite{Schwartz_1930} (P)&Unspecified&E/M&P.R.&Patent using current-driven heating of working material.\\
\hline1935&\cite{Schwartzkopf_1935} (P)&Various suggested&Either&Static (1)&First patent to suggest materials with different $T_C$; discussion of deleterious effects of thermal hysteresis.\\
\hline1940&\cite{Wehe_1940} (P)&Fe-Ni&Perm.&P.R. + Static&Used light to heat Ni-Fe alloys, driving either a Curie wheel or inducing a current in a coil.\\
\hline1945&\cite{Hindle_1945} (P)&Fe-Ni&Either&P.R.&First patent to suggest the use of waste heat.\\
\hline1948&\cite{Brillouin_1948} (A)&N/A&Perm.&Static (1)&Thermodynamic analyses of induction-based TMG: proposed Ericsson cycle.\\
\hline1959&\cite{Elliott_1959} (A)&Gd&Perm.&Static (1)&Design only: direct (induction-based) conversion of electricity based on 20 {$^\circ$}C heat source and Alnico XII permanent, fixed magnet.\\
\hline1964&\cite{Resler_1964} (A)&Fe ferrofluid (FF)&E/M&P.R.&Thermodynamic analysis of a FF circulation-based TMG without regeneration.  Here, rotation is circulation of the FF caused by the presence of the high field region.\\
\hline1967&\cite{Resler_1967} (A)&Fe ferrofluid&E/M&P.R.&Thermodynamic analysis of a FF circulation-based static TMG with regeneration.\\
\hline1969&\cite{Voort_1969} (A)&Fe ferrofluid&E/M&P.R.&Proposed modulation of applied magnetic field in FF circulation-based TMG to achieve Carnot efficiency.\\
\hline1969&\cite{Merkl_1969} (P)&Fe-Ni&Perm.&P.R.&Multiple stators in a ring structure.\\
\hline1970&\cite{Kemenczky_1970a,Kemenczky_1970b} (P)&Fe-Ni&Perm.&P.R.&Ornament, using heat to drive motion.\\
\hline1972&\cite{Murakami_1972} (A)&Unstated ($T_C\sim$50\degree C)&E/M&P.R.&Single-pole wheel with low output and efficiency; postulated use of more suitable working materials.\\
\hline1973&\cite{Pirc_1973} (P)&Gd&Perm.&P.R.&First Gd-based engine patent, using rotation of thin Gd strips.\\
\hline1977&\cite{ohkoshi_1977} (A)&NdCo$_5$&Perm.&Static (2)&First spin reorientation-based approach; induction using Pretzel-like geometry.\\
\hline1984&\cite{Katayama_1984} (P)&Soft ferrite&Perm.&P.R.&Proposed use of counter-torque from an external generator to provide smooth rotation.\\
\hline1984&\cite{Kirol_1984} (A)&Gd&Perm.&Static (1)&Thermodynamic analysis: demonstrated that thermal regeneration would improve TMG efficiency.  Power density also calculated for shunts made of Fe and Ho$_{69}$Fe$_{31}$.\\
\hline1988&\cite{Vollers_1988} (P)&Ni-Fe (example)&Perm.&P.R.&Use of solar heat or flame heat to generate motion via pyromagnetic effect.\\
\hline1988&\cite{Solomon_1988} (A)&Gd&E/M&Static (1)&Thermodynamic analysis; showed that cycling the magnetic field improves efficiency of generator.\\
\hline1989&\cite{Solomon_1989} (A)&Y$_2$(Fe,Co)$_{17}$&E/M&Static (1)&Thermodynamic analysis of a static generator with regeneration and a compositionally graded working material.\\
\hline2006&\cite{Takahashi_2006} (A)&Fe$_{54}$Ni$_{36}$Cr$_{10}$&Perm.&P.R.&3-pole, 5~W output device operating with $\Delta T_\textrm{res}$=84~K.\\
\hline2007&\cite{Ujihara2007a} (A)&Gd&Perm.&Linear&127~\textmu{}m-thickness Gd film attached to a spring; inferred potential electrical output (via piezoelectric conversion) of 1.85-3.61~mW~cm$^{-3}$.\\
\hline2008&\cite{Kitanovski_2008} (R)&Not specific&Perm.&&A report which proposed the feasibility of a 1~kW, 2 Tesla TMG device operating at 5 Hz with a $\Delta T_\textrm{res}$=95~K and a hot reservoir temperature of 120\degree C.\\
\hline2011&\cite{Degen_2011} (P)&Fe$_2$P-based&Either&Static&A patent on the use of Fe$_2$P-based materials for the direct conversion of waste heat to electricity.\\
\hline2011&\cite{Srivastava_2011} (A)&Heusler&Perm.&Static (0)&Experimental study of Ni$_{45}$Co$_5$Mn$_{40}$Sn$_{10}$ as a low thermal hysteresis alloy capable of converting heat into electricity by induction.\\
\hline2012&\cite{Vuarnoz_2012} (A)&Gd&Perm.&&A design study, comparing the exergetic efficiency of TMG, ORC, Stirling and Rankine cycles.\\
\hline2014&\cite{Christiaanse2014} (A)&(Mn,Fe)$_2$(P,As)&Perm.&Static (2)&Used 4 (Mn,Fe)$_2$(P,As) compositions in a induction-based TMG with a regenerative Ericsson cycle.\\
\hline2014&\cite{Gueltig_2014} (A)&Ni-Co-Mn-In (Heusler)&Perm.&Linear&Use of a free-standing Ni$_{50.4}$Co$_{3.7}$Mn$_{32.8}$In$_{13.1}$ Heusler alloy film, oscillating at 200 Hz, to generate electricity directly.  Average power density $\sim$1.6~\textmu{}W~cm$^{-3}$ when the periodic temperature variation in the TMG is 10~K.\\
\hline2015&\cite{CCChen2015} (A)&Gd&Perm.&Linear&Self-oscillation of a Gd cantilever between a room temperature NdFeB magnet and an ice/water container, using a piezoelectric layer to generate electricity.\\
\hline2016&\cite{wetzlar} (A)&Gd,NdCo$_5$&N/A&N/A&Modeling of a thermomagnetic cycle, showing that spin reorientation transition in NdCo$_5$ with $\Delta T_\textrm{res}$=31~K could give rise to an order of magnitude greater work density than Gd operating with $\Delta T_\textrm{res}$=5~K.\\
\hline2017&\cite{Chun_2017} (A)&Gd&Perm.&Linear&Motion of Gd film converted into electricity via coupling to a piezoelectric cantilever.\\
\hline2017&\cite{Gueltig_2017} (A)&Ni-Mn-Ga&Perm.&Linear&Self-actuating 5 $\mu$m Ni-Mn-Ga film on CuZn oscillating towards/away from a heated magnet; 84 Hz resonant operation with average power density $\sim$118 mW~cm$^{-3}$ when the periodic temperature variation in the TMG is 3~K.\\
\hline2015&\cite{Coray_2015,swiss_blue} (A)&Gd&Perm.&P.R.&Swiss Blue Energy builds a 1kW TMG system based on the Curie ring principle, with $\Delta T_\textrm{res}$=50~K.\\
\hline2018&\cite{Vida_2018} (P)&Gd&Perm.&P.R.&Patent of the (Swiss Blue Energy) device to generate circular motion from an annular ring of Gd successively warmed and cooled in the vicinity of multiple magnets.\\
\hline2019&\cite{Deepak_2019} (A)&(MnNiSi)$_{0.7}$ --- (Fe$_{2}$Ge)$_{0.3}$&Perm.&Linear&Passive device to generate current from bulk shape memory alloys, using active heating and cooling, with an output power density of 0.6~\textmu{}W~cm$^{-3}$.\\
\hline2019&\cite{Waske2019} (A)&La(Fe,Co,Si)$_{13}$&Perm.&Static (3)&Suggested genus classification of static, induction-based prototypes; demonstrated a design in 3rd genus, having 2 permanent magnets arranged to maximise power output and minimise hysteresis effects.\\
\hline2021&\cite{Ahmim_2021a} (A)&La(Fe,Co,Si)$_{13}$&Perm.&Linear&Self-oscillating device.  Active material is mounted on a cantilever, to which piezoelectric plates are also attached.  Average power density $\sim$6.8~\textmu{}W~cm$^{-3}$ and positive net power output when $\Delta T_\textrm{res}$=38~K.\\
\hline2021&\cite{Dzekan2021} (A)&La(Fe,Co,Si)$_{13}$&Perm.&Static (3)&Demonstrated that TMG efficiency is higher with La-Fe-Co-Si rather than Gd as the working material, due to the symmetrical $M(T)$ curve of the former.\\
\hline2021&\cite{Ahmim_2021b} (A)&La(Fe,Si)$_{13}$H&Perm.&Linear&Improved self-oscillating device, using flexing of a piezoelectric buzzer, on which the active material is mounted, to generate electricity.  Average power density $\sim$240~\textmu{}W~cm$^{-3}$ and positive (4.2~\textmu{}W) net power output when $\Delta T_\textrm{res}\sim$35~K.\\
\hline
\caption{Details of a selection of TMG prototype developments found in patents (P), in academic articles (A) or in reports (R).  We focus on: the type of magnetic working material, magnetic field source used (or proposed) and the type of design.  In the latter case, we use the TMG prototype categorization of Kitanovski~{\cite{Kitanovski_2020}}: passive rotary (P.R.), static induction or linear.  Static induction devices have a further genus class (0,1,2 or 3) given, according to the terminology developed by Waske et al.{~\cite{Waske2019}}.  The last column lists any other features of significance, including $\Delta T_\textrm{res}$, the temperature difference between the hot and cold reservoirs.  We note that most working materials are magnetically soft; the 1977 TMG prototype of Ohkoshi et al.~{\cite{ohkoshi_1977}}  based on spin reorientation was the first to use a magnetically hard (anisotropic) working material.}
\label{table:tmg_history}
\end{longtable}
\end{small}

\twocolumn

\subsection{Hard Ferromagnets In Past TMG Studies}
The role of hard ferromagnets in previous TMG research and patent literature has primarily been to provide the applied magnetic field required to convert heat into work.  In Table~\ref{table:tmg_history} we list a selection of the key advances in past TMG prototype development from the late 18th century to the present day, highlighting in particular the choice of working material, magnetic field source, and heat source (employed or anticipated, in the case of a design proposal).  Many of these are also discussed in Refs~{\cite{Kishore_2018}} and/or {\cite{Kitanovski_2020}}.  As can be seen from this table, hard ferromagnets dominate the field source column, with magnetic anisotropy in the working material only being explored in the scientific literature in the work of Ohkoshi in 1976 and 1977~\cite{ohkoshi_1976,ohkoshi_1977} and 30 years later in the work of Carman and co-workers~\cite{WetzlarThesis_2014,af,wetzlar}.  While not the focus of our work, several other trends are of note: (1) the preferential use of permanent magnets in later years, particularly after high-strength permanent magnets made of materials such as Nd-Fe-B became available; (2) that recent work has divided into (sub-watt) energy harvesters and larger, multi-watt systems; and (3) the increased study of passive (self-actuating) small-scale devices in the last 10 years.
 
\subsection{Single Domain Soft Ferromagnets, Hard Ferromagnets and Spin Reorientation}
\begin{figure*}
	\begin{subfigure}[t]{0.5\textwidth}
		\centering
		\includegraphics[width=7cm]{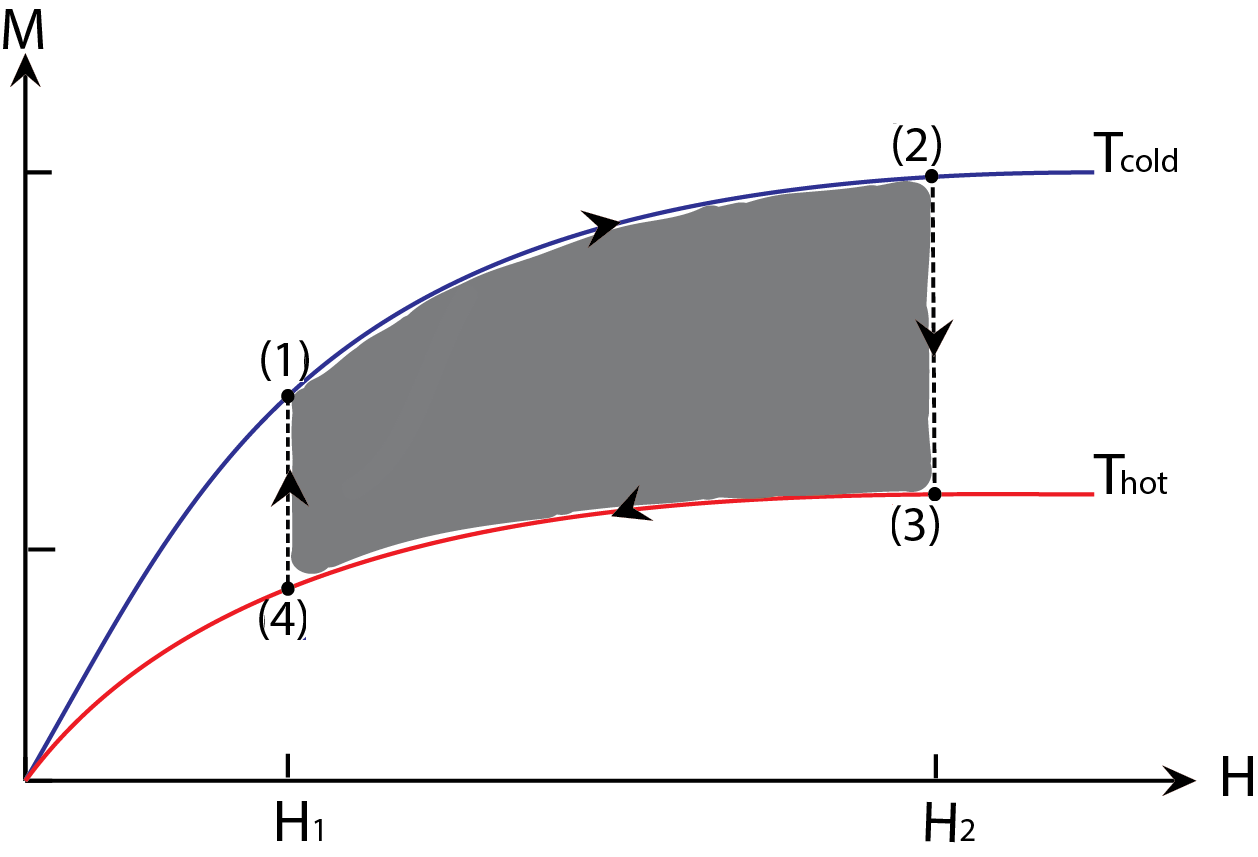}
		\caption{}
	\end{subfigure}
	\begin{subfigure}[t]{0.5\textwidth}
		\centering
		\includegraphics[width=7cm]{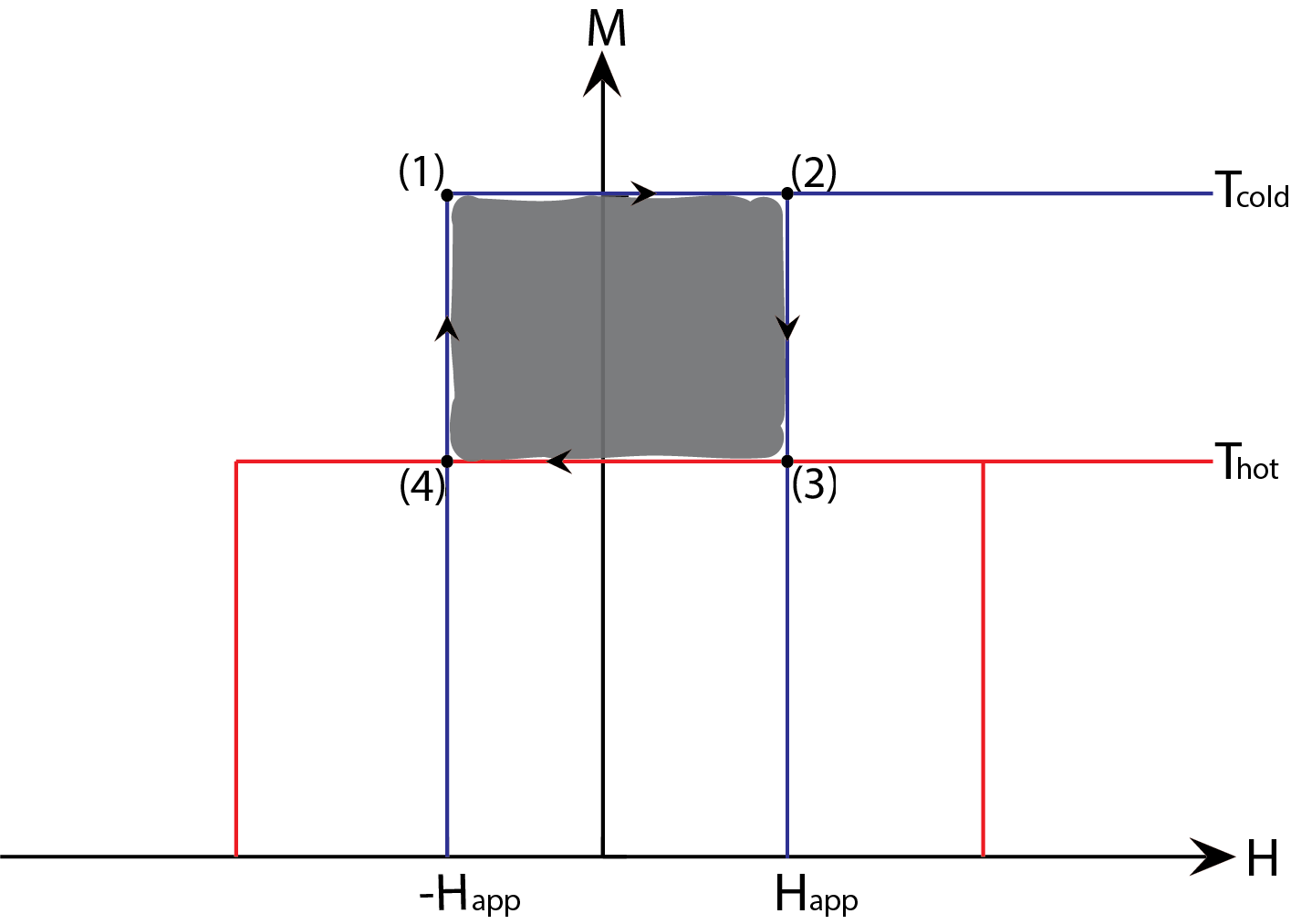}
		\caption{}
	\end{subfigure}
	\caption{Work outputs (shaded areas) of thermomagnetic cycles, operating between two temperatures, $T_{cold}$ and $T_{hot}$, and between two applied fields, $H_1$ and $H_2$.  A cycle is indicated by the numbers 1-4 in which order a positive work can be extracted from the TMG working material.  In (a) a soft ferromagnet (FM) is operated up to temperatures above its Curie point.  One quadrant of the $M-H$ is used.  In (b), a hard ferromagnet is operated below its Curie point, in positive and negative applied fields. In this case, two quadrants of the $M-H$ loop can be used.  In the schematic shown, the temperature coefficient of the coercivity is negative, which is typical for most hard magnets.  The physical example presented in Section~\ref{section:results_ferrite} is a magnetically hard ferrite with a positive temperature coefficient of coercivity.}
	\label{fig:schematic_cycles}
\end{figure*}

	This Perspective's focus on anisotropic, hard ferromagnets is strongly motivated by work conducted at UCLA by Greg Carman and coworkers~\cite{hsu,WetzlarThesis_2014,af}.  In 2011, Hsu et al. concluded that single domain gadolinium could yield a thermodynamic cycle with higher efficiency than is possible with polycrystalline, multi-domain crystals~\cite{hsu}.  The reason for the hypothesis was that a magnetically soft material with no domain walls can retain a magnetization after the removal of an applied magnetic field.  This finite remanance increases the area bounded by the hot and cold isotherms in Figure~\ref{fig:schematic_cycles}(a).  The group pursued the idea in nm-scale Gd thin films~\cite{WetzlarThesis_2014,af}.  Ultimately, the Gd nanostructures fabricated had non-uniform crystallinity, crystallographic ordering and some degree of oxidation, factors which contributed to a lack of single domain behavior~\cite{WetzlarThesis_2014,af}.  

The aforementioned previous work on single domain magnets has led us to consider the use of magnetically hard materials, which also exhibit finite magnetization at zero field, for thermal energy conversion.  We therefore present three new perspectives on the use of hard ferromagnets in a TMG cycle.  In the first two instances, we utilise the properties of established hard ferromagnets to measure or model their potential impact on a TMG cycle.  In the third case, we offer a novel route to the construction of anisotropic ferromagnets tailored to use as TMG working materials.

The first perspective is the use of a hard ferromagnet as the working material.  As demonstrated below via experimental results, the aim is to use two, rather than one, quadrant of the magnetisation vs. field ($M-H$) loop, as depicted in Figure~\ref{fig:schematic_cycles}(b).  This aim is similar to the deployment of positive and negative electric fields in a study by Basso et al. on the electrocaloric effect in the ferroelectric copolymer P(VDF-TrFE)~\cite{Basso2014}. 

The second perspective is the use of a hard ferromagnet to provide a biasing field to a magnetically soft working material.  The aim here is to achieve a higher work output than is possible without a biasing field while again utilising two quadrants in $M-H$ space.  This part of the study will be in the form of a numerical model.

The third perspective is the use of nano-structured hard magnets to provide a novel source of spin reorientation transitions (SRTs).   In addition to single domain pyromagnetism, Refs.~\cite{WetzlarThesis_2014} and \cite{af} explored SRTs as a source of pyromagnetism, based on the potential of the SRTs to generate an order of magnitude more energy with roughly comparable efficiencies \cite{wetzlar}. It was found that high-quality sample fabrication presented a significant obstacle.  Synthesized thin film, epitaxial $\rm{NdCo}_5$ samples had non-uniform crystal epitaxy which limited their utility for thermal energy applications~\cite{af,WetzlarThesis_2014}.  Our third perspective, is an alternative idea: an artificial spin-reorientation compound can potentially be formed by combining layers of two magnetic compounds with different easy axis alignments and temperature coefficients of magnetic anisotropy.  These artificial compounds may be easier to fabricate while presenting a similar pyromagnetic effect to their thin-film counterparts.  Our presentation of this idea is conceptual.

The rest of this article is organized as follows. Section~\ref{methods} details: the materials and experimental methods used our study of hard ferrites as working materials, and the computational methods used to analyze the effect of a bias magnetic field on a magnetically soft working material.   Section~\ref{results} contains the main results of those two studies and a discussion of the concept of artificial SRTs as pyromagnetic working materials.  Conclusions are drawn in Section~\ref{conclusions}.
	
\section{Methods}
\label{methods}

\subsection{Magnetically Hard Working Materials}

\paragraph{Materials} Two hexaferrite (BaFe$_{12}$O$_{19}$) samples were studied in a in-house-developed magnetometer at the Ne\'el Institute in Grenoble, France.  One is a cylindrical HF8/22 grade magnetically-isotropic ferrite measuring 3~mm in height and 3~mm in diameter, with a reported $BH_\textsubscript{max}$ of 6.5 to 6.8~kJ/m$^3$, provided by Distrelec in Switzerland.  The other is a cylindrical, anisotropic Y30BH grade, measuring 5~mm in height and 4~mm in diameter, with a reported $BH_\textsubscript{max}$ of 27.5 to 32~kJ/m$^3$, provided by AMF Magnetics in Australia.  

\paragraph{Experiments} An initial viability test was conducted for each material by measuring isothermal magnetization loops for a wide range of temperatures between -2~T and 2~T, as shown in Figure~\ref{fig:loops}.  Then, the $M-H$ loops were analyzed to select isotherms which yield significant enough changes in magnetization to generate substantial work outputs.  For the second set of tests, both materials were cycled between -0.25~T and 0.25~T, although an operational temperature range of 200~K (from $T_{\mathrm{cold}}$ = 320~K to $T_{\mathrm{hot}}$ = 520~K) was used for Y30BH, whereas a range of 320~K (from $T_{\mathrm{cold}}$ = 280~K to $T_{\mathrm{hot}}$ = 600~K) was used for HF8/22.  Each sample was loaded into the magnetometer, and an initial magnetic field of 2~Tesla was applied at $T_{\mathrm{cold}}$ in order to ensure an axial magnetization in the positive z-direction, regardless of sample orientation. Then, the field was set to -0.25~T and the loops proceeded in a clockwise fashion, as enumerated by points (1) through (4) in Figure~\ref{fig:cycles}.  Each sample was cycled through four loops.  The work output for each loop was calculated by numerically integrating the $M-H$ isotherms, as outlined in Hsu et al.~\cite{hsu}.

\subsection{Soft Ferromagnets Biased By a Hard Magnet} 

\begin{figure}
	\centering
	\includegraphics[width=\columnwidth]{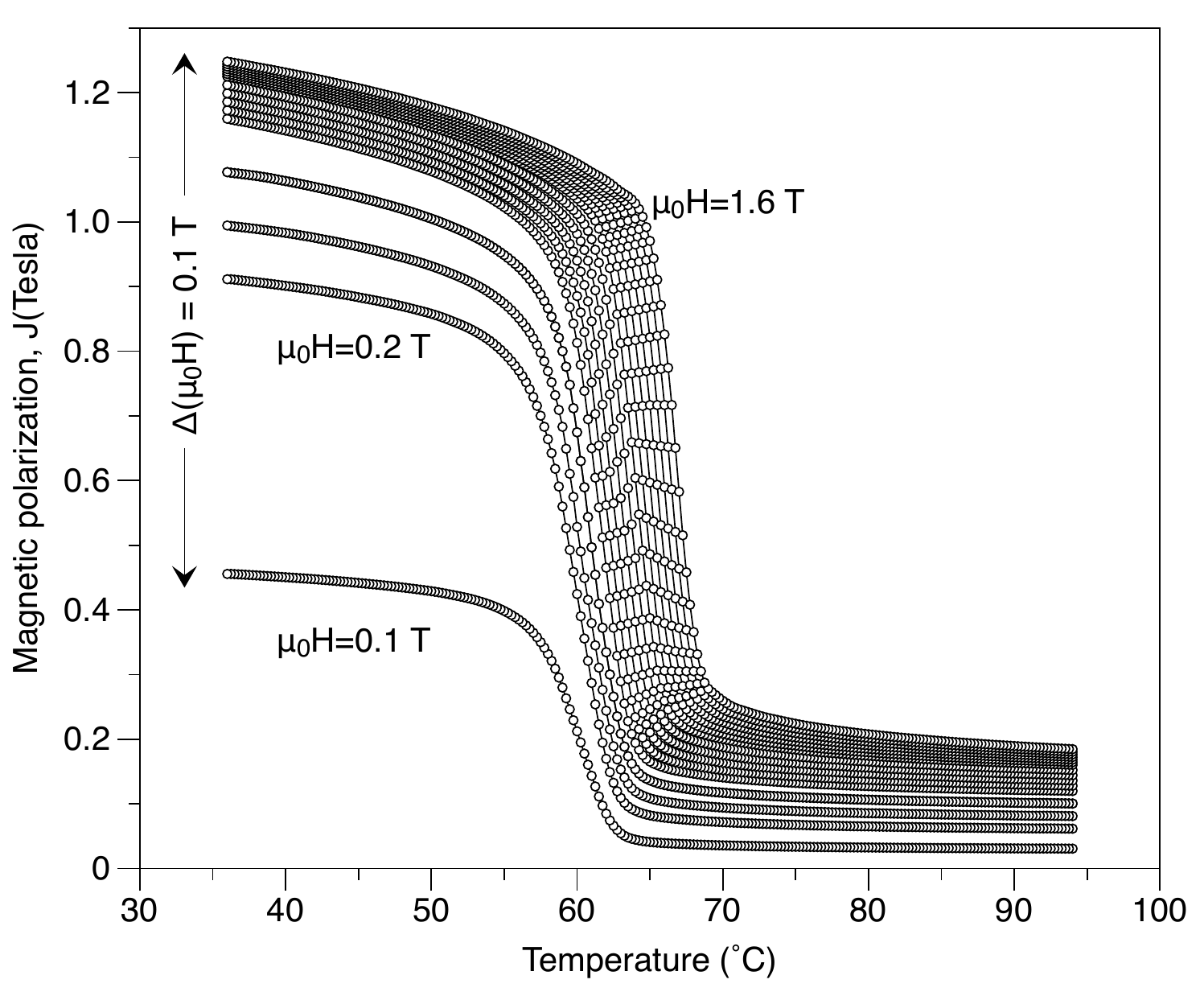}
	\caption{Iso-field magnetic polarization, $J$, for the LCFS-H sample used in Section~\ref{results:soft}.  Data were taken on cooling.  
	}
	\label{fig:mcp1625-mh}
\end{figure}

\paragraph{Materials} The interaction of magnetically soft magnetocaloric materials with a hard magnet were simulated using magnetization data from the literature.  The 2 magnetocaloric materials in the simulation are: (i) a Ce-substituted, fully hydrogenated {La(Fe,Si)$_{13}$} sample (nominal composition {La$_{0.82}$Ce$_{0.18}$Fe$_{11.8}$Si$_{1.2}$H$_{\delta}$}, henceforth LCFS-H), with first order {T$_C$} = 60\degree C; and (ii) Gd, with {T$_C$} = 25{\degree} C.   The LCFS-H material was prepared at Vacuumschmelze according to methods described elsewhere~{\cite{Katter2008}}.  Data for magnetization as a function of temperature (on cooling, from 94\degree C to 36\degree C in steps of 0.25\degree C) in fixed applied fields ($\mu_{0}H$ up to 1.6~Tesla, in steps of 0.1~Tesla) for LCFS-H were taken at Vacuumschmelze and a subset of the data is shown in Figure~{\ref{fig:mcp1625-mh}} whereas data for Gd were extracted from the study by Hsu et al.~{\cite{hsu}}. In our interaction simulations, each material was placed next to a magnetically hard NdFeB magnet, which was simulated by parameters from \textit{Radia}~\cite{radia}, a magnetostatics package for \textit{Mathematica}.  Each simulation used a temperature range of 50\degree C, centered on the Curie temperature, $T_{\rm{C}}$, of the soft material.  For LCFS-H, the temperature range was from 35\degree C to 85\degree C, and for Gd,  the temperature range was 0\degree C to 50\degree C.  A 50\degree C temperature range was chosen to enable comparison to Hsu et al.~\cite{hsu}.  In that study, a work output of 13.6~J/kg for polycrystalline Gd was found under an applied field variation from 0~T to 0.3~T and a temperature range of 0\degree C to 50\degree C.

\paragraph{Simulations} The details of the simulation are as follows.  First, we constructed the geometric constraints for the problem by defining the dimensions and relative locations for the hard and soft materials.  In both cases, we used a square face measuring 3~mm $\times$ 3~mm.  The thickness of each material was varied in order to obtain relative size effects.  Each material was magnetized through its thickness.  Three-dimensional $M(H,T)$ surface plots of magnetization as a function of temperature and magnetic field were interpolated linearly from the data for LCFS-H and Gd.  \textit{Radia} simulations produced values for the magnetic field due to the NdFeB hard magnet as a function of temperature and the amount of hard magnetic material.  The magnetic field values thus obtained were used to select the magnetization value of each soft material at every temperature from the $M(H,T)$ surface plots.
 
We conducted two sets of simulations: one for Gd with NdFeB; the other for LCFS-H with NdFeB.  In each simulation, we varied the thickness of each material to study relative size effects.  The isothermal $M-H$ curves were used to calculate work outputs.  

\section{Results and Discussion}
\label{results}

\subsection{Magnetically Hard Working Materials}
\label{section:results_ferrite}
\begin{center}
\begin{table*}
\begin{center}
\begin{tabular}{| p{0.07\linewidth} | >{\centering} p{0.12\linewidth} | >{\centering}p{0.125\linewidth}|>{\centering}p{0.08\linewidth}|>{\centering}p{0.15\linewidth}| >{\centering\arraybackslash} p{0.1\linewidth}|}
\hline
\textbf{Sample Grade} & \textbf{Field Range (T)} & \textbf{Temperature Range (K)} & \textbf{Work Output (J/kg)} & \textbf{Percentage Loss of Work Output after nth Loop}  & \textbf{Loop Orientation} \\
\hline \hline
Y30BH & -0.25 to 0.25 & 320 to 520 & -10.12 & 1st: 0.82;\newline 2nd: 0.63;\newline3rd: 0.004 & CCW \\
\hline
Y30BH & -0.25 to 0.25 & 320 to 520 & 9.63 & 1st: 4.19;\newline 2nd: -0.07;\newline3rd: 0.08 & CW \\
\hline
HF8/22 & -0.25 to 0.25 & 280 to 600 & 7.21 & 1st: 9.60;\newline 2nd: 0.56;\newline3rd: 0.23 & CW \\
\hline
HF8/22 & -0.25 to 0.50 & 280 to 600 & -16.66 & 1st: 0.70;\newline 2nd: 0.42;\newline3rd: 0.10 & CCW \\
\hline
\end{tabular}
\caption{Summary of magnetization data, experimental parameters, and calculated work outputs for the two hard ferrite samples studied here.  Cycles were performed in a clockwise (CW) or counterclockwise (CCW) loop in $M-H$ space. We note the change in the work output apparent after the first cycle (expressed as a percentage change relative to the first loop) and that this change is greatest for CW cycles.}
\label{table:wo}
\end{center}
\end{table*}
\end{center}
	
\begin{figure*}
	\begin{subfigure}{0.5\textwidth}
		\centering
		\includegraphics[width=8cm]{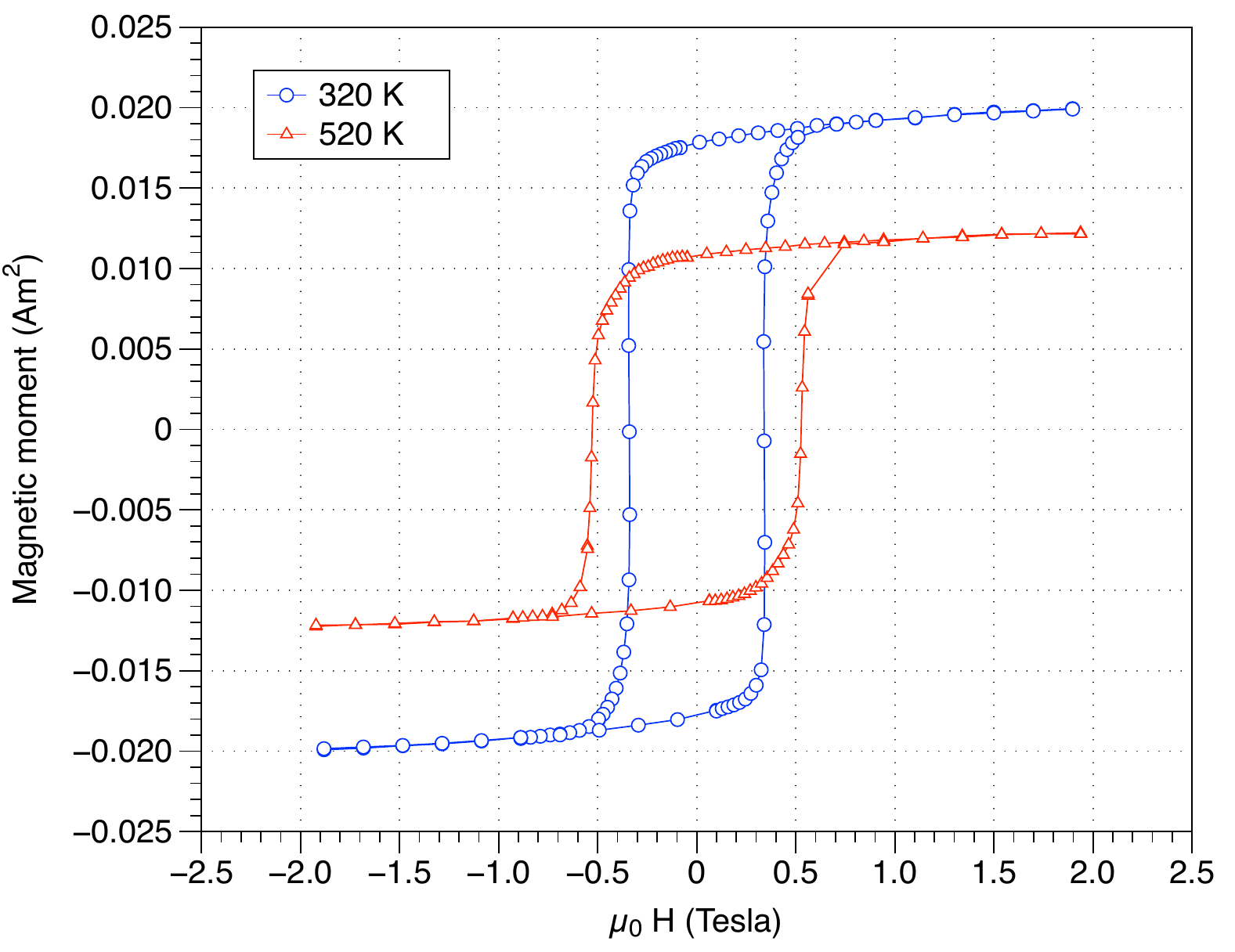}
		\caption{Y30BH grade ferrite.}
	\end{subfigure}
	\begin{subfigure}{0.5\textwidth}
		\centering
		\includegraphics[width=8cm]{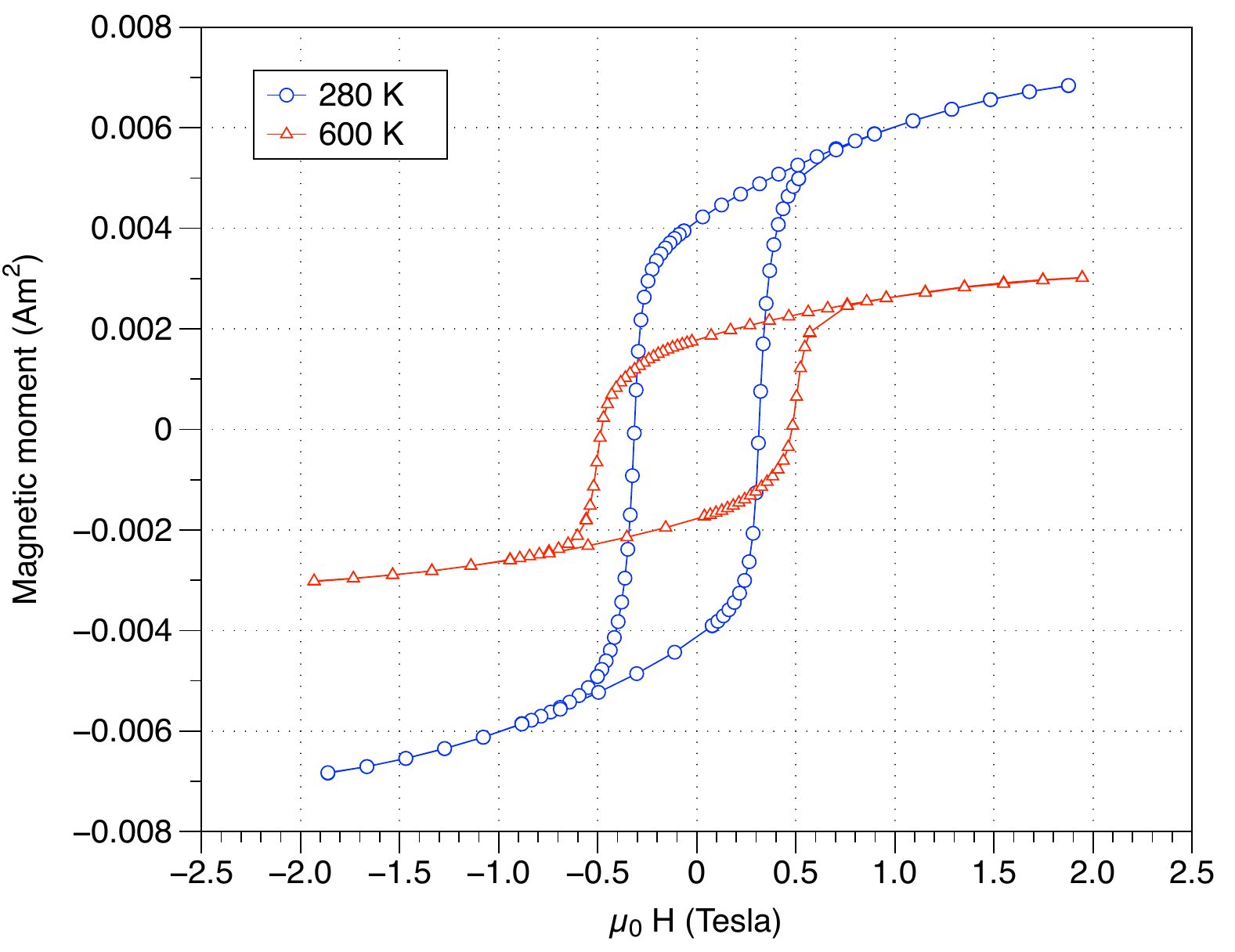}
		\caption{HF8/22 grade ferrite.}
	\end{subfigure}
	\caption{Isothermal magnetization loops for the two ferrite samples studied in this work.  The Y30BH grade sample is magnetically anisotropic while the HF8/22 grade ferrite is isotropic.  Both loops are corrected for demagnetizing fields.  We note the similarity of these loop shapes, particularly for the Y30BH sample, to the desired schematic loops in Figure~\ref{fig:schematic_cycles}(b).}
	\label{fig:loops}
\end{figure*}

The first novel perspective in this article is in considering ferrites as magnetically hard TMG working materials.  The purpose is to increase the efficiency associated with the TMG cycle while bringing down the cost of the functional material.  The efficiency $\eta$ of a TMG cycle was given in Equation~\ref{efficiency_equation}.  The integral $\oint \! M \mathrm{d}H$ is the work output from one complete cycle around a $M-H$ loop.  In this context, the reasons for selecting ferrites (apart from their cost) are twofold.  Firstly, hard ferrites have a positive temperature coefficient of coercivity, meaning the coercivity increases as temperature increases.  Using hard ferrites therefore avoids the risk of thermal demagnetization under magnetic fields close to the low-temperature coercive field.  Magnetization loops for the two ferrites studied in this work are shown in Figure~\ref{fig:loops}.   Secondly, magnetically hard materials have magnetization loops with two positive magnetization quadrants in the magnetisation vs applied field ($M-H$) plane.  This offers a new perspective on TMG: by opening up the possibility to construct two-quadrant thermomagnetic generation cycles, wherein the absolute value of the applied magnetic field is small.  Low magnetic fields give small contributions to the magnetically-driven entropy change $\Delta{S}$~\cite{hsu}.

\begin{figure*}
	\begin{subfigure}[t]{0.5\textwidth}
		\centering
		\includegraphics[width=8cm]{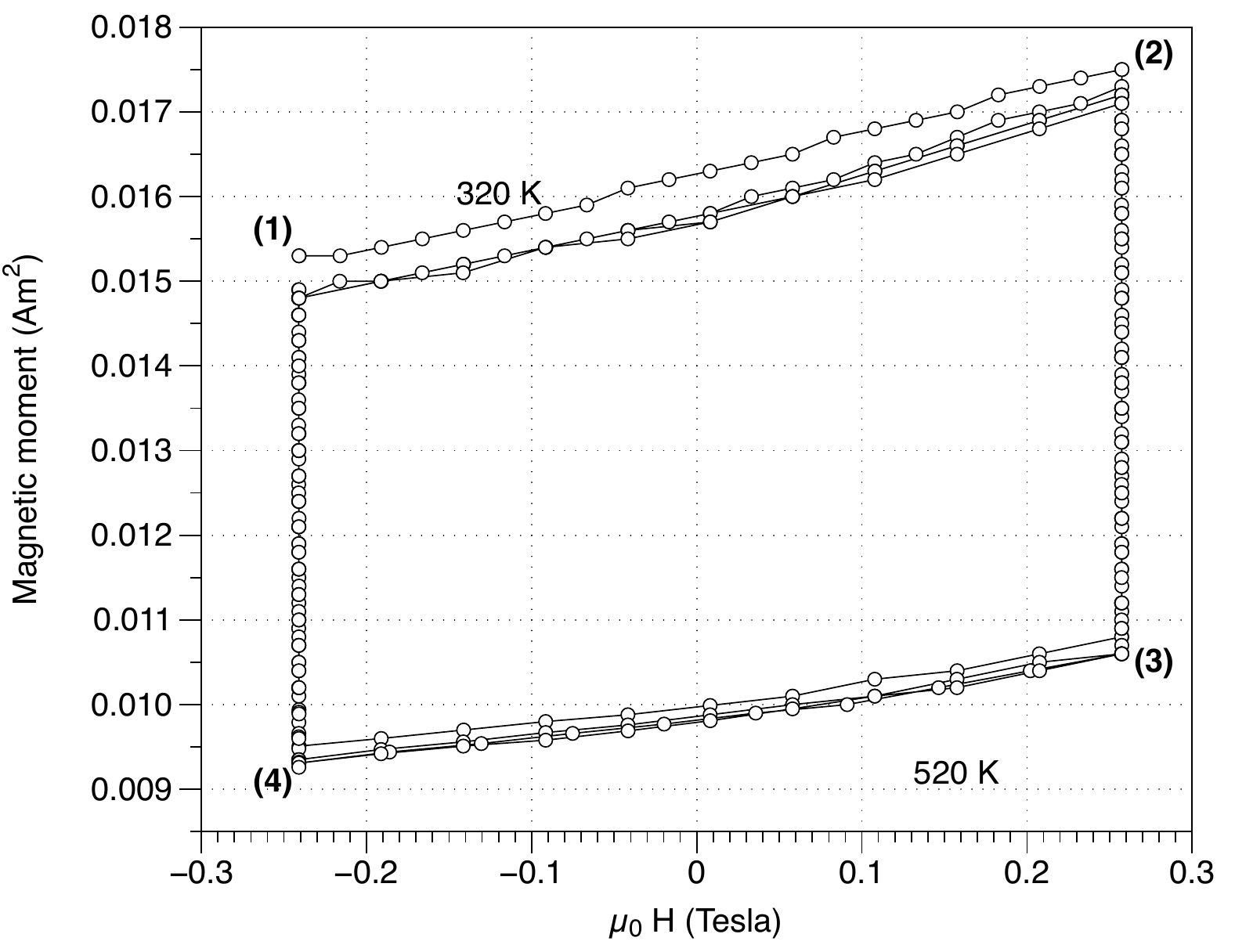}
		\caption{Y30BH: 320 K to 520 K.}
	\end{subfigure}
	\begin{subfigure}[t]{0.5\textwidth}
		\centering
		\includegraphics[width=8cm]{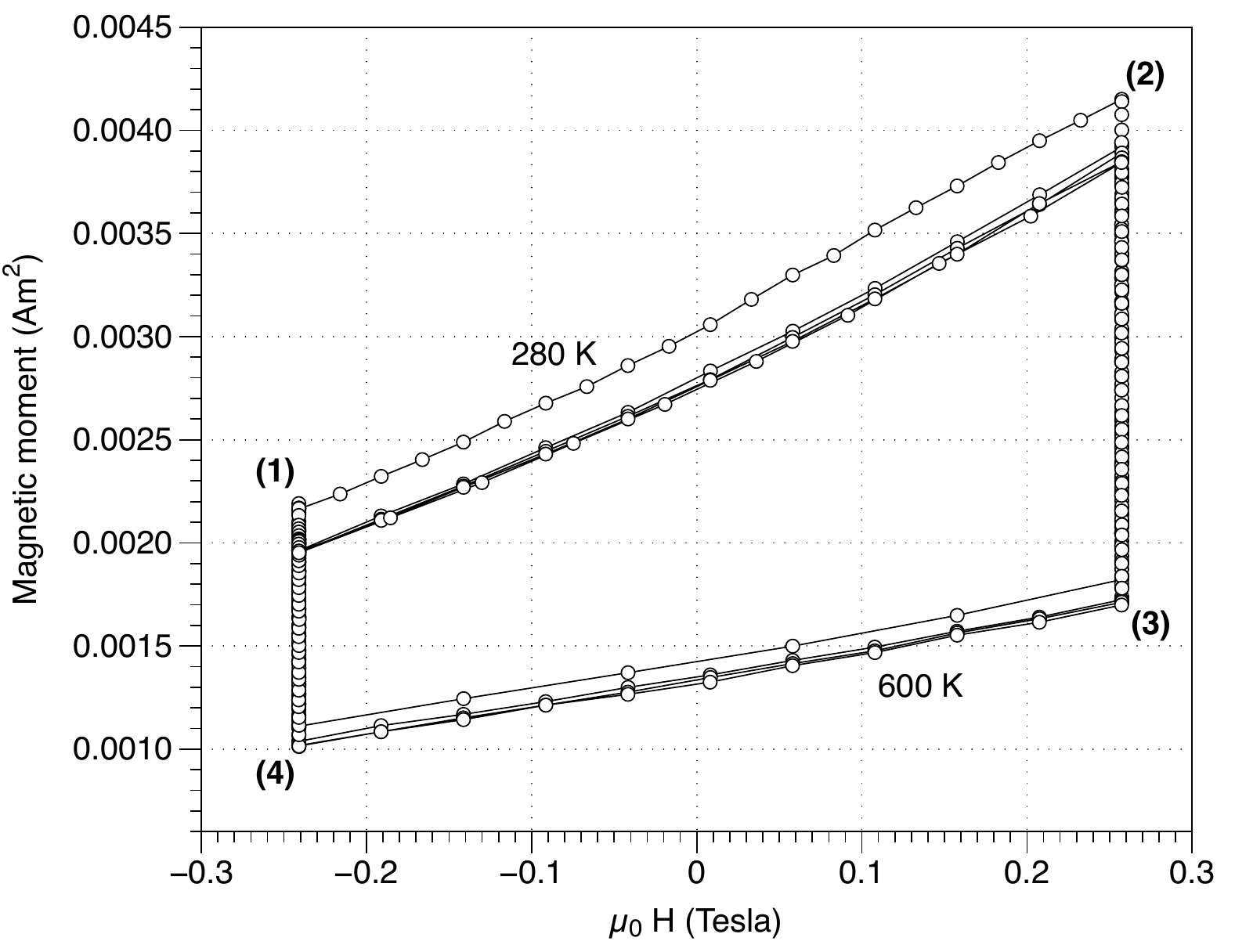}
		\caption{HF8/22 ferrite: 280 K to 600 K.}
	\end{subfigure}
	\caption{Thermomagnetic generation cycles for two hard ferrite samples for applied fields between -0.25 T and 0.25 T.  Loops proceed clockwise from (1) to (4) on each graph. The average work outputs (excluding the first cycle) for the cycles shown are (a) 9.8 J/kg and (b) 6.9 J/kg.}
	\label{fig:cycles}
\end{figure*}

The results of example magnetization measurements taken by thermomagnetic cycling are shown in Figure~\ref{fig:cycles}.  These cycles consisted of two isotherms (in which the sample temperature was maintained by the magnetometer) and two iso-field steps.  Previous work has demonstrated that the major  loops shown in Figure~\ref{fig:loops} possess a superposition of a (reversible) magnetocaloric effect and an irreversible entropy production~\cite{Basso_2010a}.  We present cycles for which the applied magnetic field does not substantially exceed  the coercive field and therefore assume that the material stays within the reversible regime.  

The data shown in Figure~{\ref{fig:cycles}} were taken for clockwise (CW) cycles (positive work output); data were also taken for counterclockwise (CCW) cycles (negative work output, see Table~{\ref{table:wo}}).  The CW minor loops shown in Figure~{\ref{fig:cycles}} possess opposite concavity to that exhibited by the counter-clockwise major loops shown in Figure~{\ref{fig:loops}}.  The CCW loops (not shown) exhibited the same concavity at the major loops in Figure~{\ref{fig:loops}}.  In both sets of loops, the increasing field leg of the first cycle (points 1 to 2) is noticeably higher than in the remaining cycles.  This is an artifact of the experimental procedure.  When the samples were set into the magnetometer, an initial magnetic field of +2.0 T was applied to ensure magnetization in the positive direction, regardless of sample orientation. This imparts an additional magnetic remanence to the sample, which lingers until thermal demagnetization (points 2 to 3 on the labelled graphs).

Table~\ref{table:wo} contains the work outputs associated with clockwise (CW) and counterclockwise (CCW) cycles performed on both samples.  In addition, loops under cycling between magnetic field values of -0.25~T and 0.25~T, an additional loop was performed on the HF8/22 sample between -0.25~T and +0.5~T.  Furthermore, the relative loss of work output upon mutiple cycles is shown as an additional column.

The highest performing material tested was Y30BH grade ferrite, with a work output of 9.8~J/kg under cycling between magnetic field values of -0.25~T and 0.25~T and a temperature range from 320~K to 520~K.  This work output compares well with that of polycrystalline Gd, which provides 13.6~J/kg under cycling between field values of 0~T to 0.3~T and a temperature range from 273~K to 323~K (or 0\degree C to 50\degree C). Although the work outputs are comparable, Gd performs far better under realistic TMG circumstances, which call for operating frequencies on the order of 0.5~Hz to 2~Hz in medium to large-scale devices and which therefore utilise smaller temperature ranges.  Hence, further work to optimize hard magnetic materials is required to bring about a new class of magnetically hard, high performance TMG materials.

From Table~{\ref{table:wo}} we also see that there is a reduction of work output upon multiple cycling. It is noticeable from the 1st to the 2nd cycle, i.e. after the first cycle, in both samples, and by the 2nd cycle the work output is more stable.  The initial loss of work output is associated with the additional magnetic remanence mentioned above, and from exposing the sample to a magnetic field in the opposite direction to its delivered magnetization.  We see that the loss is greatest for clockwise magnetization loops.  However, the physics of the change of work output on cycling deserves further exploration, as relaxation of a magnetic loop towards the anhysteretic curve is a potential risk, if the chosen magnetic cycle lies far away from the anhysteretic curve.  Indeed, while the magnetic loops made in our experiments are well controlled and consist of isotherms and iso-field steps, conditions in a device might deviate from this and the impact of such deviations is unknown.  Lastly, the physical mechanism underlying the work loss may depend on the choice of hard magnet material, which might affect the viability of alternatives to magnetically hard ferrites.
\\
\subsection{Soft Ferromagnets Biased By a Hard Magnet} 
\label{results:soft}

\begin{figure}
	\centering
	\includegraphics[width=\columnwidth]{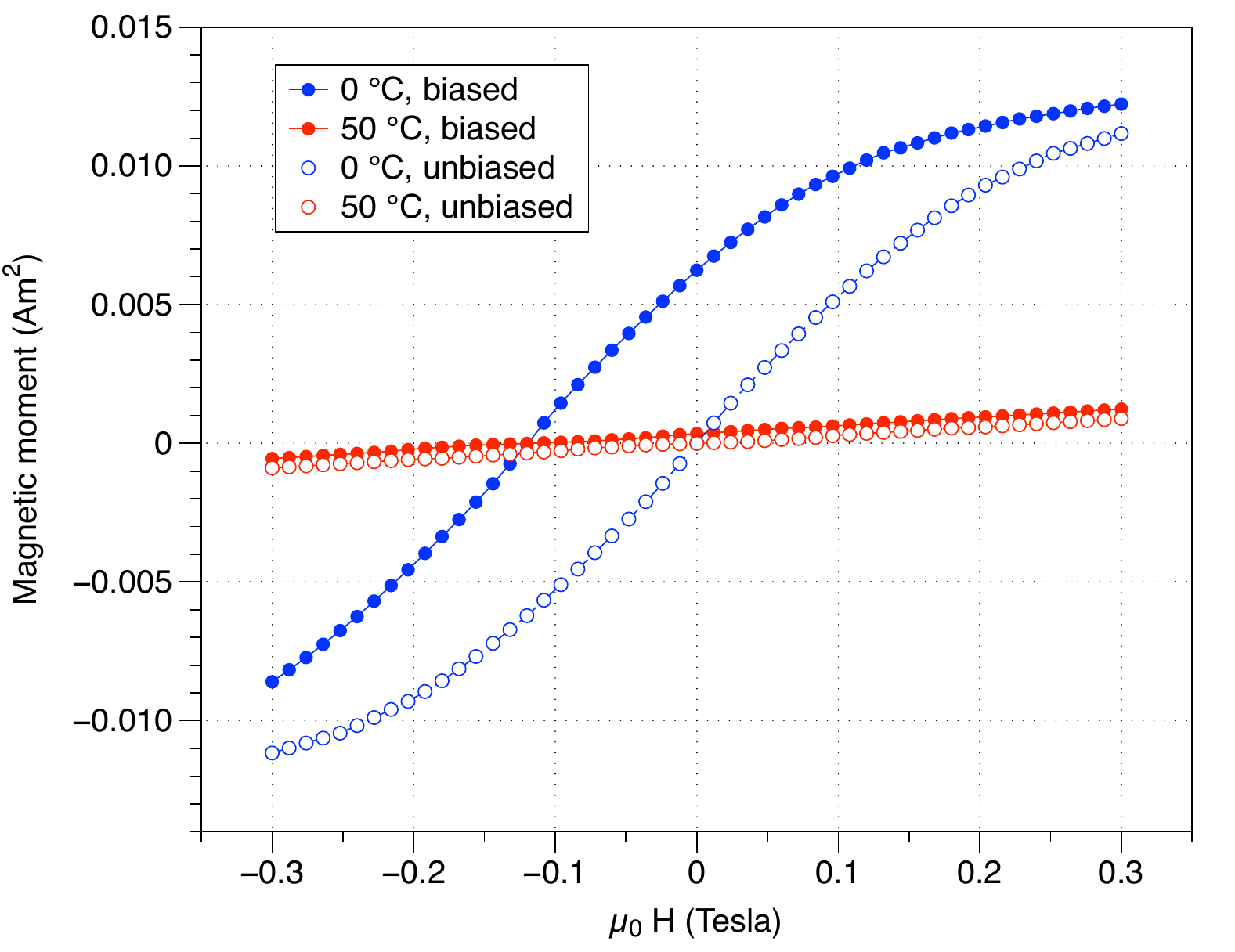}
	\caption{Simulated magnetisation loops for gadolinium under no magnetic biasing (open circles, dashed lines) and biasing from a NdFeB magnet (solid circles, solid lines). The extra work output available when biasing is present is given by the area between the two blue curves, for moment values greater than zero.}
	\label{fig:gd}
\end{figure}

Since the work output, $W$, for any material in a thermomagnetic power generation cycle can be given by Equation~\ref{work_eq}, we may alter the construction of the magnetization loops in order to maximize the work output.  This can potentially be achieved by using a hard ferromagnet to apply a local biasing magnetic field to a magnetically soft material.  The effect of bias electric or magnetic fields have previously been considered in a few caloric materials.  The effect of an internal bias electric field has been examined in an analytical Landau model of an electrocaloric material~\cite{Ma2018}.  The effect of a biasing external magnetic field has been considered for direct (induction-based) TMG cycles on a Heusler alloy with a first order magnetostructural transition, with the motivation of converting as much magnetic work into electricity as possible~\cite{Song2013}.  In this Perspective, our purpose is different from either example: if the magnetic field from the hard magnet is opposite to the magnetic field applied to the entire system, then both positive and negative applied (external) magnetic fields can be used in the TMG cycle.  The results of simulations using two such systems are detailed below.

\begin{figure}
	\centering
	\includegraphics[width=\columnwidth]{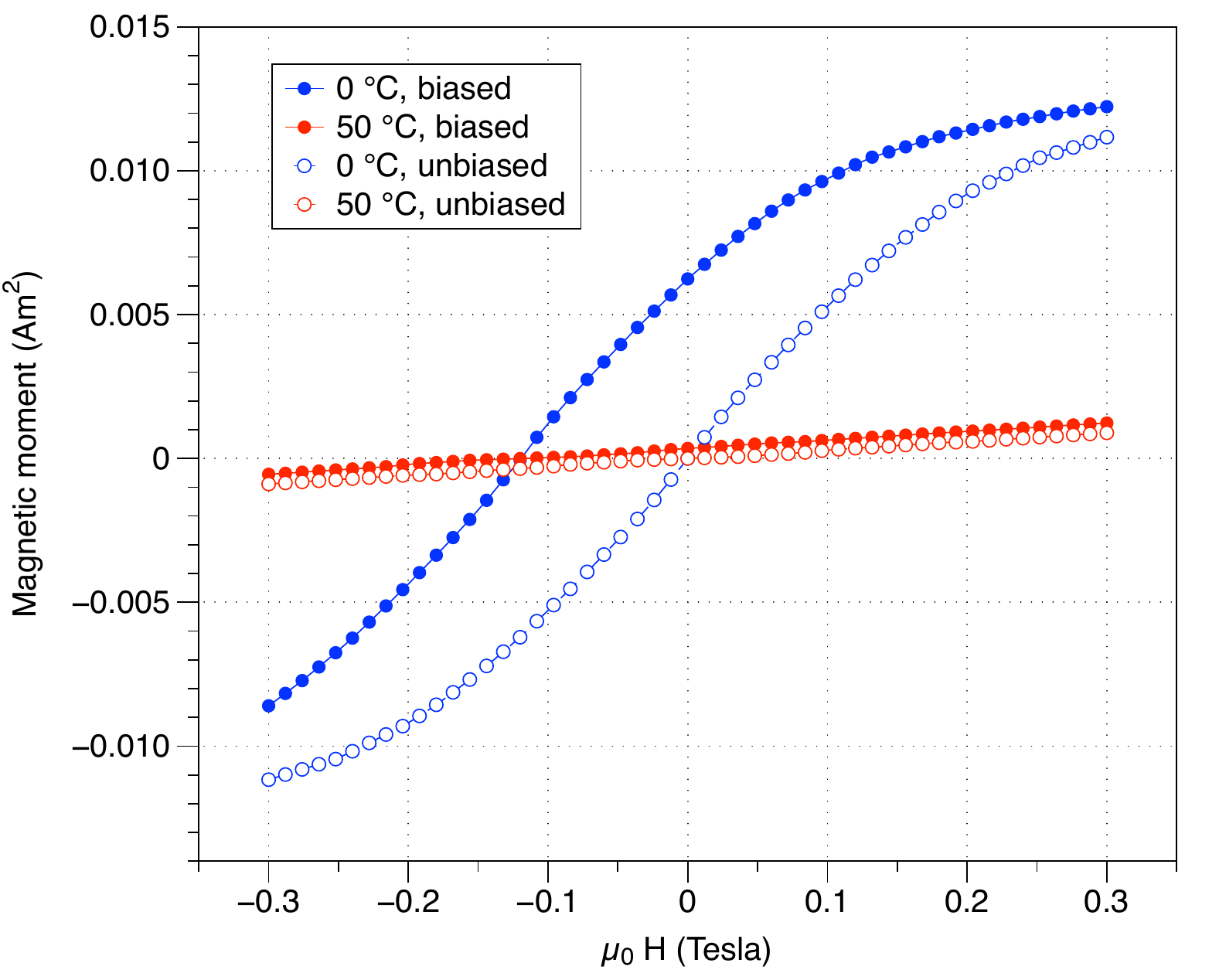}
	\caption{Examples of simulated magnetisation loops for LCFS-H under no magnetic field biasing (open circles, dashed lines) and under biasing from a NdFeB magnet (solid circles, solid lines). The extra work output available when biasing is present is given by the area between the two blue curves, for moment values greater than zero.}
	\label{fig:lfs}
\end{figure}

Magnetization loops for Gd and LFS under both unbiased and biased conditions can be seen in Figures \ref{fig:gd} and \ref{fig:lfs}, respectively.  To calculate the maximum available work for the materials when they are magnetically biased, a thermodynamic cycle between the largest negative field value that still yields a positive work output and 0.3~T is considered.  Also, both TMG materials are cycled through a temperature range of 50\degree C, centered about the Curie temperature.  The benchmark values with which our results are compared are the computed work outputs for unbiased materials under an external applied field which varies from 0 T to 0.3~T when the materials are cycled over a 50\degree C temperature range about their Curie points.  For Gd, the benchmark work output thus computed is 13.6~J/kg, and for LCFS-H, the benchmark value computed is 20.4~J/kg.

\begin{figure}
	\centering
	\includegraphics[width=\columnwidth]{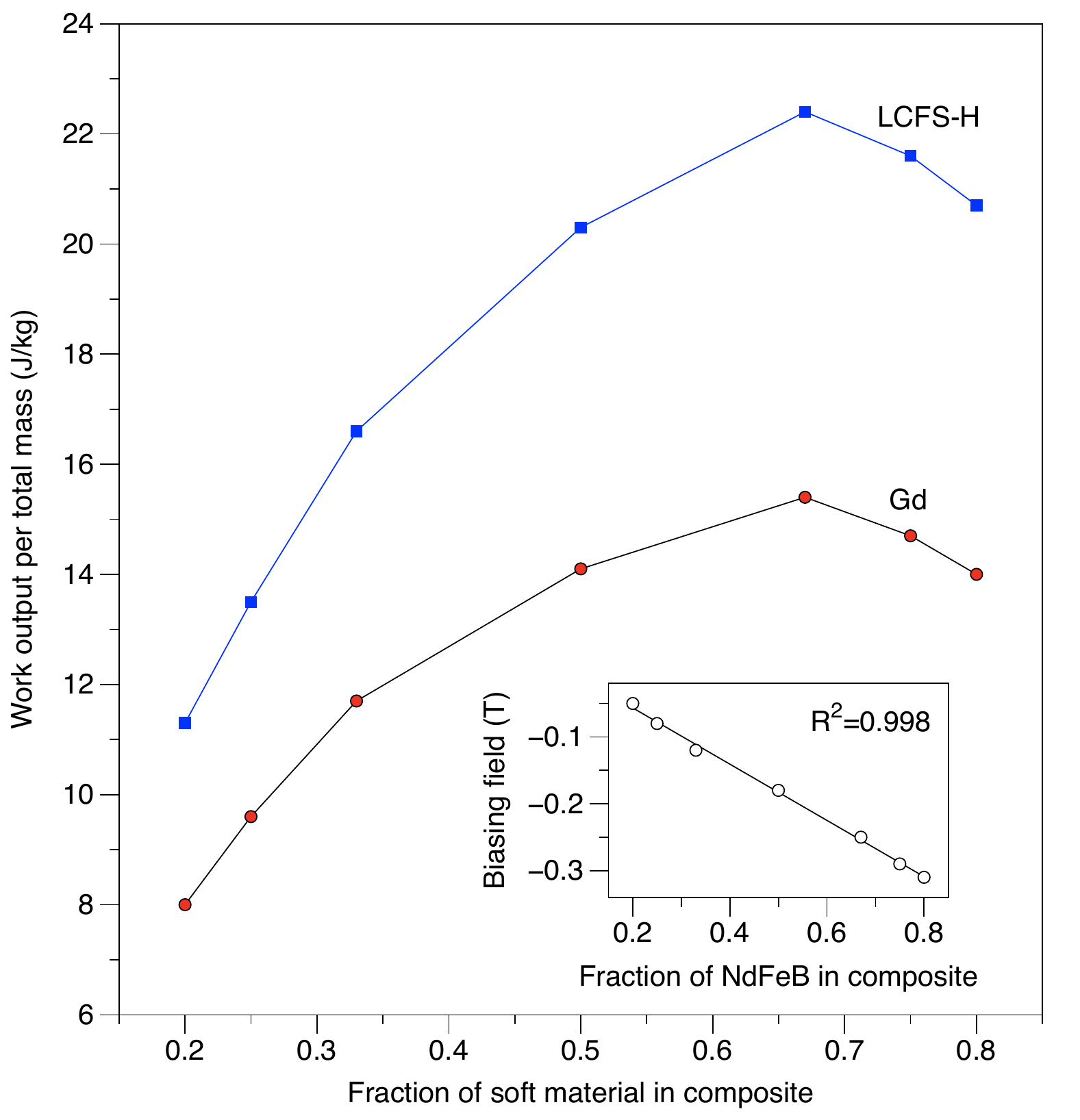}
	\caption{Relative size effects as they impact the work output per total system mass and (inset) the biasing field provided by the NdFeB portion of the composite.}
	\label{fig:comp}
\end{figure}

In both Figure~\ref{fig:gd} and Figure~\ref{fig:lfs}, we see that the impact of the bias field is considerable; the external applied field can now be negative as well as positive, yielding additional utilization of the external magnetic field, just as in the hard ferrite example, but via a different mechanism.  In addition, the relative size effects from varying the amounts of hard and soft magnetic materials in the system have a significant impact on the results.  The maximal work output per total system mass is achieved for a combination of 67\% soft TMG magnet and 33\% hard biasing magnet for both of the soft materials tested, as presented in Figure~\ref{fig:comp}(a).  At this material composition ratio, the biased cycle presents an improvement in the work output compared to the benchmark of each material: a 13\% improvement in the case of Gd and 9.8\% for LCFS-H.  These modest numbers are dramatically improved when one no longer accounts for the mass of the hard magnet; when computing work outputs per mass of soft material alone, the work outputs for composites comprising 67\% soft ferromagnet are 22.6 J/kg (Gd) and 33.7 J/kg (LCFS-H).  The latter values correspond to a 65\% increase over the unbiased cycle.  Further numerical details for composites with different soft:hard magnet ratios are given in~\cite{tantillo_thesis}.

The maximum negative field that can be used while obtaining only positive contributions to the work output is equal to the magnitude of the biasing field from the hard magnet.  The latter is plotted as a function of hard magnet fraction in the inset to Figure~\ref{fig:comp}.  It is also worth considering how finite size effects affect the hard magnet.  It is intuitive that a lower fraction of NdFeB in the composite would lead to a smaller biasing field, but the data presented in the inset to Figure~\ref{fig:comp} confirm the linear increase of biasing field with increasing hard magnet composition.  

\subsection{Spin Reorientation Transition TMG}
\label{section:SRT_discussion}
To this point of our Perspective, the only mechanism for TMG materials that has been discussed involves a thermally-driven change in the magnitude magnetization vector.  Although thermally-driven changes in the magnitude of the magnetization vector can yield significant TMG work output, it is also worth considering other mechanisms for magnetization changes.  Spin reorientation transitions occur in materials which experience a temperature-driven change in the direction of the magnetization vector from one crystallographic axis to another.  By changing the projection of the magnetization in a certain direction, it is possible to drive a change in magnetic flux, which can then be harvested to do work.  A brief review of embodiments of this idea, some promising results in the literature on SRT TMG, and some initial concepts for further development, are presented below. 

The first published spin-reorientation TMG design was created by Ohkoshi et al. in 1977~\cite{ohkoshi_1977} after an initial conceptual proposal in the previous year~\cite{ohkoshi_1976}.  Their magnetic ``pretzel'' consisted of an SRT material (NdCo$_5$), a magnet for the applied field (SmCo$_5$), iron yokes to direct the magnetic flux, and pickup coils to convert flux changes directly into electrical energy.  The work output achieved by the pretzel design, in terms of the amount of SRT material, was 1.5 J/cm\textsuperscript{3}, which corresponds to 177~J/kg of NdCo$_5$.  This work output is significantly larger than the values found in previous sections for Gd or LCFS-H, albeit for cycles with different input thermal and magnetic parameters. 

Further calculations and analysis by Wetzlar et al.~\cite{wetzlar} confirmed that the available work associated with a TMG cycle for NdCo$_5$ operated around its SRT could be 4 times larger than that associated with single domain Gd operated around its Curie point, albeit with a different operating temperature ranges in each case (30~K for NdCo$_5$ and 5~K for Gd).  We note that single-domain Gd should have a larger work output than poly-domain Gd due to the finite remanence mentioned above, and so the potential for SRTs to provide large work outputs is considerable.

However, there is a drawback to SRTs in single-phase materials: a sharp SRT is normally only observed in highly textured or single crystalline samples.  This can place a restriction on sample morphologies and the scalability of the final device.  For example, as noted earlier, NdCo$_5$ films can be prone to oxidation~\cite{af}.  Therefore, to complete this portion of our Perspective, we briefly outline a concept, the details of which we leave to a separate article~\cite{basso_2022}: namely, that SRT could also be accomplished using artificial magnetic composites, which could be significantly easier to find or fabricate than single-crystal SRT materials.  

To start, we state a simplified form of the magnetic free energy for a uniaxial magnetic system:
\begin{equation}
G = K_{A}\sin{^2\theta_A} - \mu_0 \mathbf{M}\cdot\mathbf{H},
\end{equation}
where $K_A$ is the temperature-dependent anisotropy constant and $\theta_A$ is the angle of the easy-axis. (Note that only the first term of the anisotropy energy, which can be expanded in even powers of sine, is considered.)  If thin layers of materials A and B with different anisotropy constants ($K_A$ and $K_B$) and volume fractions ($v_A$  and $v_B$) are arranged so that the easy axes are perpendicular to one another, the free (anisotropy) energy becomes
\begin{equation}
f = v_{A}K_{A}\sin{^2\theta_A} + v_{B}K_{B}\sin{^2(\frac{\pi}{2}-\theta_B)},
\end{equation}
where $\theta_B$ is measured with respect to the easy axis of material A.  If the constants $K_A$ and $K_B$ have different temperature dependencies, there will be a critical temperature at which the magnetizations of the two materials change from being perpendicular to aligning along a single axis.  At this temperature, there is an artificial spin-reorientation transition, which can be used to generate energy from magnetic flux in the same way as with single-crystal materials.  Further details will be elaborated in a standalone work~\cite{basso_2022}.

TMG based on artificial SRT is distinct from Curie point TMG in several ways.  Firstly, while $\Delta{S}$, the magnetic field-driven isothermal entropy change depends explicitly on the applied magnetic field in pyromagnets with second order Curie transitions, $\Delta{S}$ is independent of field for SRT-TMG, provided the field is large enough to induce a spin-reorientation transition~\cite{ilyn}.   Secondly, if an artificial SRT can be constructed from building blocks which do not themselves contain a SRT, there should be good opportunities for tailoring the final SRT properties by adjusting the thermomagnetic properties of the materials A and B and their relative dimensions.  Lastly, shape-dependent demagnetizing factors, magnetoelastic stresses in the material, and the magnitude of the applied magnetic field all play a role in the value of SRT work outputs, making the power output and efficiency of a SRT-based TMG highly tunable~\cite{wetzlar}.

\section{Conclusions}
\label{conclusions}
We have used experimental data, physical models and conceptual arguments to provide three new perspectives on the use of hard magnetic materials in TMG cycles.  Firstly, we have demonstrated that known hard ferromagnetic working material can provide work outputs comparable to that of Gd, albeit using operating temperature spans which are larger than desired.  The hard magnet is able to make use of two quadrants in the magnetization-applied field plane, meaning that an applied field may be reversed and extract further energy.  This could help to reduce the magnetic field required in TMG, if a hard magnet can be designed in which the coercivity is more thermally stable than the magnetization.  Secondly, we have modeled the effect of an applied bias magnetic field on a soft ferromagnetic working material, showing again that a two-quadrant $M-H$ loop can be used, reducing the magnetic field need.  Lastly, we have outlined the potential for artificial spin reorientation transitions to be explored for TMG applications, by tailoring the properties of hard magnetic films with competing thermal coefficients of magnetic anisotropy.

\section*{Acknowledgements}
We are grateful to Yves Deschanels for assistance with magnetometry measurements and Dominique Givord, Laurent Ranno, Morgan Almanza and Fr\'ed\'eric Mazaleyrat for enlightening discussions.  This work has benefited from the support of the project HiPerTherMag ANR-18-CE05-0019 of the French National Research Agency (ANR). Support for this project was provided by a PSC-CUNY Award, jointly funded by The Professional Staff Congress and The City University of New York (TRADA-49-452). A.N.T. received financial support from The Grace Spruch ’47 Physics Fund at Brooklyn College. K.G.S. would like to acknowledge the Université Grenoble Alpes for sponsoring his invited teacher researcher position during 2017.

\section*{References}

\bibliography{tmg_bibfile}

\end{document}